\documentclass[aps,prb,twocolumn,groupedaddress,amsmath,amssymb,showpacs]{revtex4}

\usepackage{graphicx}
\usepackage{hyperref}
\usepackage{pifont}
\begin{document}
%Title of paper
\title{Dimensionality driven charge density wave instability in TiS$_2$}
\author{Kapildeb Dolui and Stefano Sanvito}
\affiliation{School of Physics, AMBER and CRANN Institute, Trinity College, Dublin 2, Ireland}
\date{\today}
%*********************************************************************
%ABSTRACT
%*********************************************************************
\begin{abstract}
Density functional theory and density functional perturbation theory are used to investigate the 
electronic and vibrational properties of TiS$_2$. Within the local density approximation the material 
is a semi-metal both in the bulk and in the monolayer form. Most interestingly we observe a Kohn anomaly in 
the bulk phonon dispersion, which turns into a charge density wave instability when TiS$_2$ is thinned 
to less than four monolayers. Such charge density wave phase can be tuned by compressive strain, 
which appears to be the control parameter of the instability.
\end{abstract}
\maketitle
%*********************************************************************
% INTRODUCTION
%*********************************************************************
%\section{Introduction}
{\it Introduction.} Titanium disulfide (TiS$_2$), a prototypical layered transition metal dichalcogenide 
(LTMD), consists of covalently bonded Ti and S atoms arranged in two-dimensional hexagonal planes 
(monolayers). In the bulk form monolayers are stacked together by weak van der Waals forces. Many 
conflicting results exist in literature on the electronic structure of bulk TiS$_2$, namely it is still debated 
whether the material is metallic, semi-metallic or semiconducting. Several experimental works report 
TiS$_2$ to be a semiconductor with a bandgap ranging from 0.05 to 2.5~eV. These observations are 
supported by angle-resolved photo-emission spectra~\cite{PRB_1980_21_615}, optical 
absorption~\cite{JPChS_1965_26_1445}, Hall~\cite{JPCssp_1984_17_2713, JPC_1977_10_705} and 
transport measurements~\cite{JPC_1983_16_393}. At the same time a second set of experiments 
places TiS$_2$ among metals and semi-metals with an indirect bandgap overlap ranging from 0.2 to 
1.5~eV (between the $\Gamma$ and the L point in the Brillouin zone). This experimental evidence is based 
on extensive resistivity measurements~\cite{PRL_1972_29_163}, infrared 
reflectance~\cite{PRB_1973_7_3859} and X-ray emission and absorption~\cite{PRB_1973_8_3576}.

TiS$_2$ has drawn considerable attention over the past four decades because of the variety of its 
potential applications. For instance, alkali atoms and organic molecules can be intercalated easily into 
TiS$_2$ for applications related to both light-weight and high-energy-density 
batteries~\cite{AdvP_1987_36_1, Sc_1976_192_1126}. 
In particular, it has been demonstrated that for bulk TiS$_2$ the intercalation changes the 
conductivity~\cite{AdvP_1989_38_565}. A second interesting aspect is that both the electronic structure 
and the electron transport properties of TiS$_2$ can be modulated by external pressure. Many 
experimental~\cite{JPCssp_1984_17_2713, JAP_2011_109_053717} as well as 
theoretical~\cite{JPCssp_1985_18_1595, JPCM_2011_23_055401, SSSc_2010_12_1786} studies 
describe pressure induced phase transitions in TiS$_2$. For instance, the transition from a semiconductor 
to a semi-metal phase has been supported by Hall measurements on TiS$_2$ single crystals 
under a 4~GPa pressure~\cite{JPCssp_1984_17_2713}. Such evidence highlights the strong
coupling between the electronic properties and the TiS$_2$ structure, a feature shared by many
LTMDs.

In fact, the LTMD family in general exhibits strong electron-phonon coupling~\cite{AdvP_1987_36_1}. Often 
this results into ground states displaying a macroscopic order. For instance bulk NbSe$_2$ and TaSe$_2$ 
are notoriously known to exist in both a superconducting and a charge-density wave (CDW) 
state~\cite{NJP_2008_10_125027}. Recently a significant experimental effort has been devoted to
produce ultra-thin layers of such class of materials either by synthesis~\cite{NatNano_2012_7_699} 
or by exfoliation~\cite{JCScience}. The appeal of reduced dimension LTMDs is that their 
electronic and vibrational properties are strongly modified when the thickness reduces down to ultra-thin 
layers, resulting in further potential for future applications. For example, photo-luminescence emerges due 
to an indirect to direct bandgap change when the thickness of MoS$_2$ is reduced from bulk to 
monolayer~\cite{NL_2010_10_1271}. In contrast, the TiSe$_2$ phonon spectrum softens with decreasing 
the layers thickness~\cite{NL_2012_12_5941}.

In this work, we study the changes in the electronic and vibrational properties of the TiS$_2$, when the 
material is thinned from bulk to the monolayer limit. In particular we demonstrate that, if the material
is metallic, a CDW phase will emerge when the number of layers is reduced below four. Intriguingly the 
phase transition can be driven by compressive strain.

%*********************************************************************
%METHODOLOGY
%*********************************************************************
%\section{Methodology}
{\it Methodology}.
The electronic and vibrational properties of bulk and monolayer TiS$_2$ have been investigated by using 
respectively {\it ab-initio} density functional theory~\cite{PR_136_B864, PR_140_A1133} and density 
functional perturbation theory~\cite{RMP_73_515}. We have considered the local density 
approximation (LDA)~\cite{PRL_1980_45} of the exchange and correlation functional, as implemented in 
the {\it Quantum Espresso} package~\cite{QE}, except for a set of calculations (see later) where the 
LDA+$U$~scheme\cite{LDAU}, still within {\it Quantum Espresso}, has been employed. Ultra-soft pseudo-potentials 
describe the core electrons of all the atomic species. The electronic wavefunction is expanded using plane waves 
up to a cutoff energy of 70~Ry. Calculations for bulk TiS$_2$ are performed for a periodic structure with the  
P$\overline{3}m$1 space group. In contrast, for TiS$_2$ monolayers a 12~\AA~vacuum region is inserted along 
the non-periodic direction ($z$-axis). Geometries are relaxed by conjugate gradient, where both the atomic positions 
and the cell parameters are allowed to relax until the forces on each atom are less than 0.005~eV/\AA. The electronic 
integrations are carried out by using a 16$\times$16$\times$10 (20$\times$20$\times$1) Monkhorst-Pack $k$-grid for 
bulk (monolayer) TiS$_2$ and a Hermite Gaussian smearing of 0.01~Ry is used for all the calculations. 
In computing the phonon spectrum the dynamical matrix is evaluated over a 4$\times$4$\times$2 
(4$\times$4$\times$1) phonon-momentum grid and it is interpolated throughout the Brillouin 
zone in order to plot the bulk (monolayer) phonon bandstructure. 

\begin{figure}[ht]
\center
\includegraphics[width=7.0cm,clip=true]{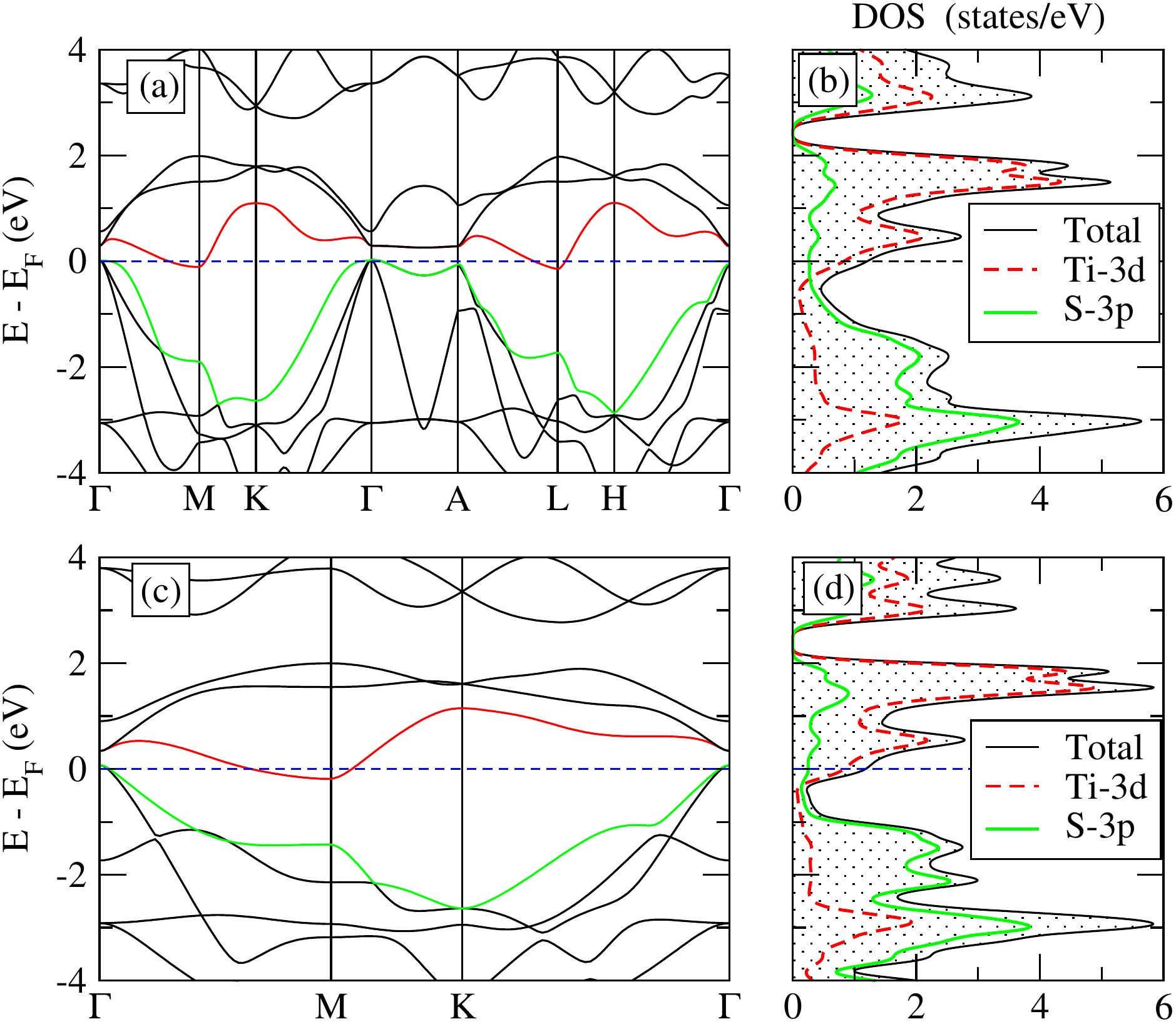}
\caption[]{(Color online) The LDA-bandstructure of bulk (a) and monolayer (c) TiS$_2$. The conduction band (red line) 
and valence band (green line) overlap between the $\Gamma$ and the L (M) points for bulk (monolayer) TiS$_2$. 
The density of states (black line) is projected over the Ti 3$d$ (red dotted line) and S 3$p$ (green line) orbitals for 
bulk (b) and monolayer (d). The blue dashed line indicates the Fermi level, $E_\mathrm{F}$. The electronic states in 
the DOS are Gaussian broadened by 0.01~Ry.}
\label{fig:band_DOS}
\end{figure}

%*
%******************************************************************
%RESULT AND DISCUSSION
%******************************************************************
%\section{Results and discussion}
%**********************************************************************
{\it Results and discussion.} We begin by calculating the electronic properties of bulk TiS$_2$. The optimized 
bulk cell parameters are $a=b=3.31$~\AA, $c=5.45$~\AA, in good agreement with the previous theoretical 
(LDA) values~\cite{JPCM_2011_23_055401} of $a=b=3.312$~\AA, $c=5.449$~\AA\ and with the experimental 
ones~\cite{PRB_57_5106} of $a=b=3.407$~{\AA}, $c=5.695$~{\AA}. The Ti-S bond length is found to be 
2.38~{\AA}, again in close agreement with the experimental value~\cite{AdvP_1969_18_193} of 2.32~{\AA} and 
with a previous theoretical (LDA) estimate~\cite{SSSc_2010_12_1786} of 2.383~{\AA}. The calculated 
energy bandstructure along the high symmetry lines in the bulk TiS$_2$ (P$\overline{3}m$1) Brilloiun zone is shown 
in Fig.~\ref{fig:band_DOS}(a). Clearly bulk TiS$_2$ is predicted to be a semi-metal with an indirect band-overlap of 
0.16~eV (negative bandgap). The valence band maximum is found to be located at the $\Gamma$ point, 
while the conduction band minimum is at L. The corresponding partial density of states (DOS) projected over 
the different atomic species shows that the valence and the conduction band originate respectively 
from the S 3$p$ and the Ti 3$d$ orbitals, although there is a good degree of hybridization, see 
Fig.~\ref{fig:band_DOS}(b). This is in good agreement with previous calculations~\cite{JAP_2011_109_053717}.

Next we move to study the electronic properties of TiS$_2$ in its monolayer form [see figures \ref{fig:band_DOS}(c)
and \ref{fig:band_DOS}(d)]. Our optimized lattice constant, $a=b$, is now 3.32~{\AA}. Such value slightly underestimates
the experimentally observed one~\cite{ACIE_2011_50_11093} of 3.40~{\AA}, as expected from the covalent nature 
of the Ti-S bond and the general tendency of the LDA to overbind. Our calculations show that the electronic property, 
i.e. the bandgap, does not change significantly when the thickness decreases from bulk to a single monolayer. 
Our computed LDA negative bandgap of -0.25 eV is in good agreement with the previous theoretical 
calculations~\cite{JPCC_116_8983}. Also in TiS$_2$ monolayers, like in the bulk, the conduction and valence bands 
are respectively derived from Ti $3d$ and S $3p$ orbitals [Fig.~\ref{fig:band_DOS}(d)].

\begin{figure}[ht]
\center
\includegraphics[width=8.0cm,clip=true]{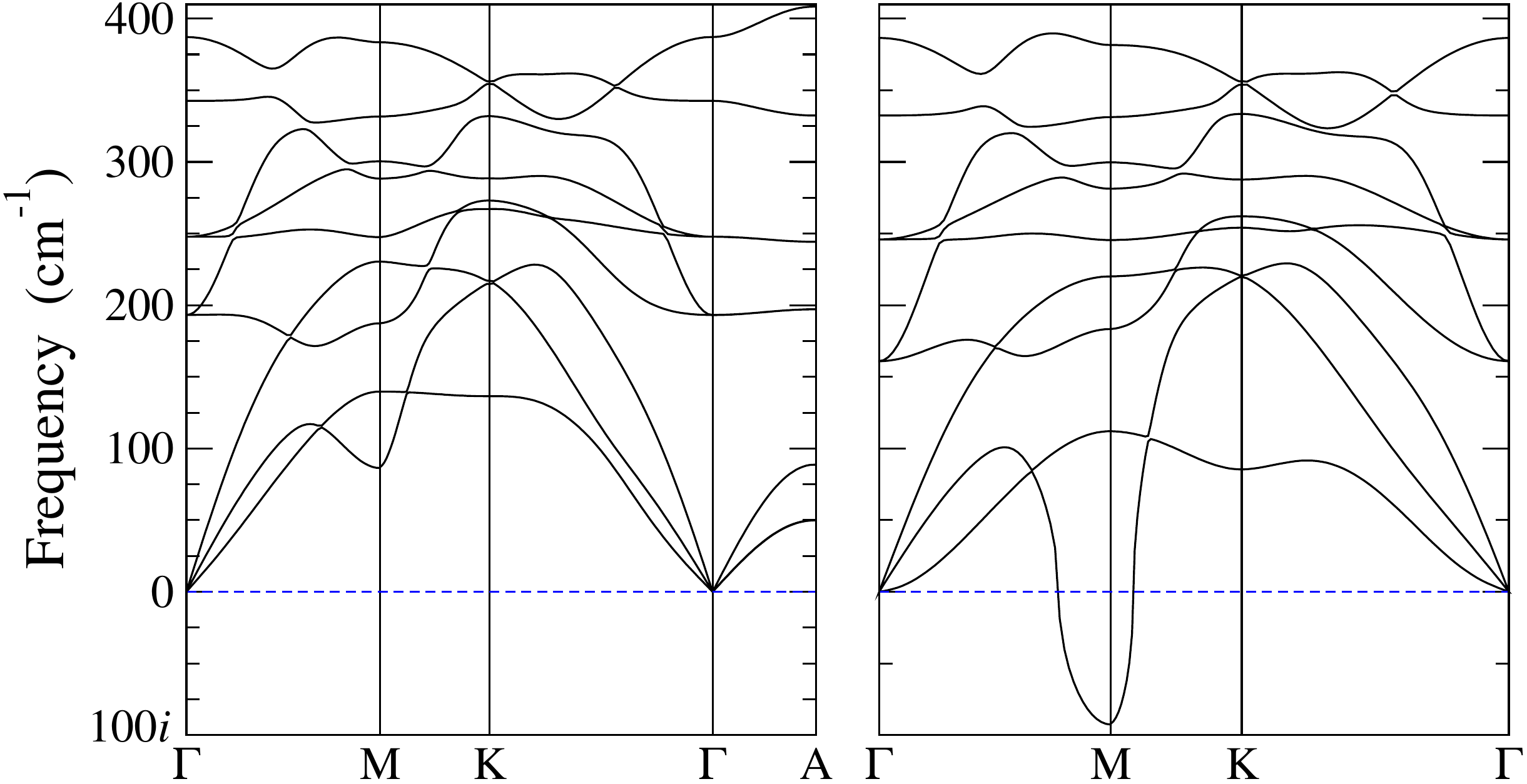}
\caption[]{The phonon dispersion of bulk (a) and monolayer (b) TiS$_2$. Note that the calculated imaginary 
frequencies are shown as negative roots of the square phonon frequencies. Lines are guides to the eye.}
\label{fig:PH_pristine}
\end{figure}

\begin{figure}[ht]
\center
\includegraphics[width=8.0cm,clip=true]{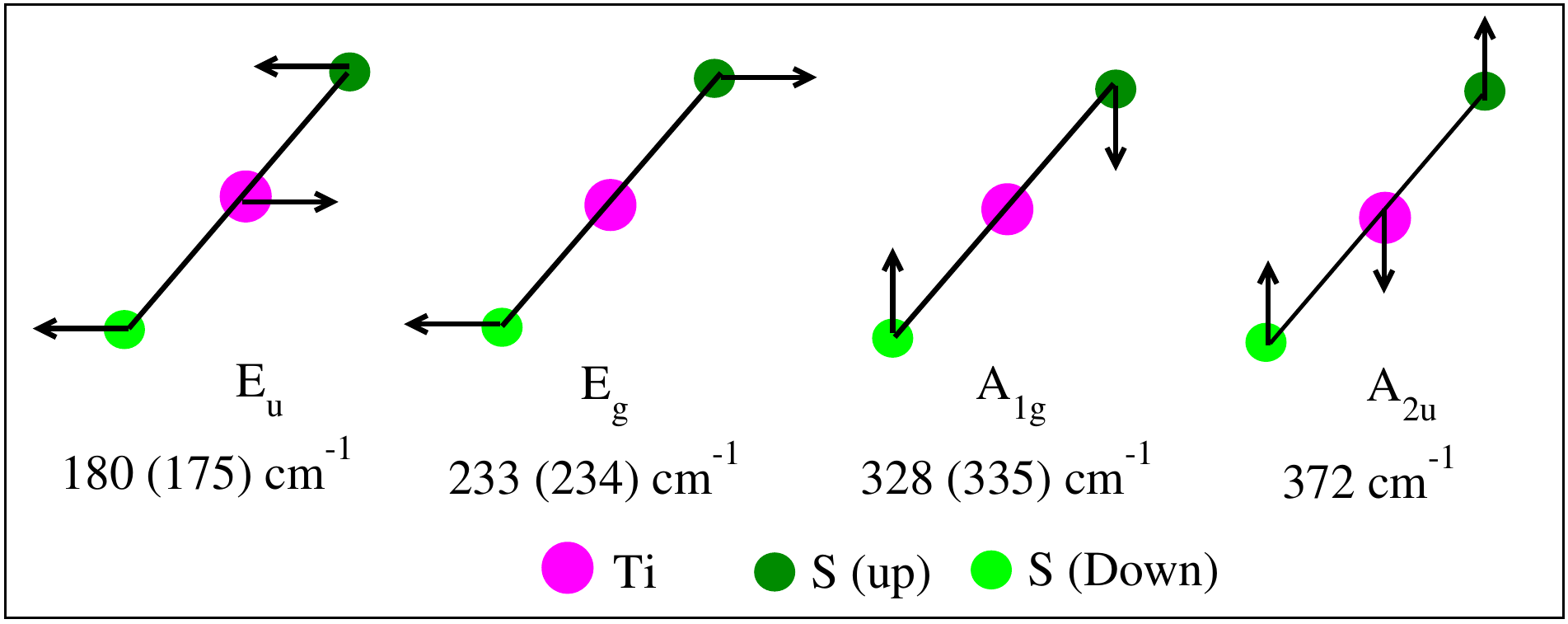}
\caption[]{(Color online) The atomic displacements of the $\Gamma$-centered optical 
phonon modes of bulk TiS$_2$. Frequencies are reported from our calculations while the corresponding 
experimental values~\cite{PRB_1992_45_14347} are in bracket. S (up) and S (down) refer to S atoms respectively
on the upper and lower plane relative to the Ti one.}
\label{fig:modes}
\end{figure}

We now investigate the effects of the dimensionality on the phonon properties of TiS$_2$ by starting from its
bulk form, whose phonon dispersion is plotted in Fig.~\ref{fig:PH_pristine}(a) along the $\Gamma$-M-K-$\Gamma$-A 
$q$-path. In order to facilitate a comparison with experiments we recall that the irreducible
representation of the zone-center phonon modes reads
\begin{eqnarray}
\Gamma = 2A_{2u}(\mathrm{IR}) + 2E_u(\mathrm{IR}) + A_{1g}(\mathrm{R}) + E_g(\mathrm{R}),
\end{eqnarray}
where R and IR denote respectively Raman and infrared active modes. The corresponding atomic displacements 
of the optical modes at the $\Gamma$-point are shown in Fig.~\ref{fig:modes}. Notably the phonon frequencies of 
a number of modes have been measured experimentally by vibrational spectroscopy, such as Raman, infrared 
and neutron scattering~\cite{PRB_1992_45_14347, PRB_1986_33_4317}. These values are also reported in 
Fig.~\ref{fig:modes}, demonstrating a good agreement with our calculated ones. In addition, our phonon 
dispersion agrees quite well with previous theoretical calculations based either on
empirical-valence-force-field methods~\cite{PRB_1992_45_14347} or on state of the art first principle density 
functional perturbation theory~\cite{JPCM_2011_23_055401}. Interestingly, we observe the signature of a 
soft-phonon mode, generally called Kohn anomaly, at the M point in the Brillouin zone. This signals an incipient 
structural instability, which is yet not fully realized in the bulk.

\begin{figure}[ht]
\center
\includegraphics[width=8.0cm,clip=true]{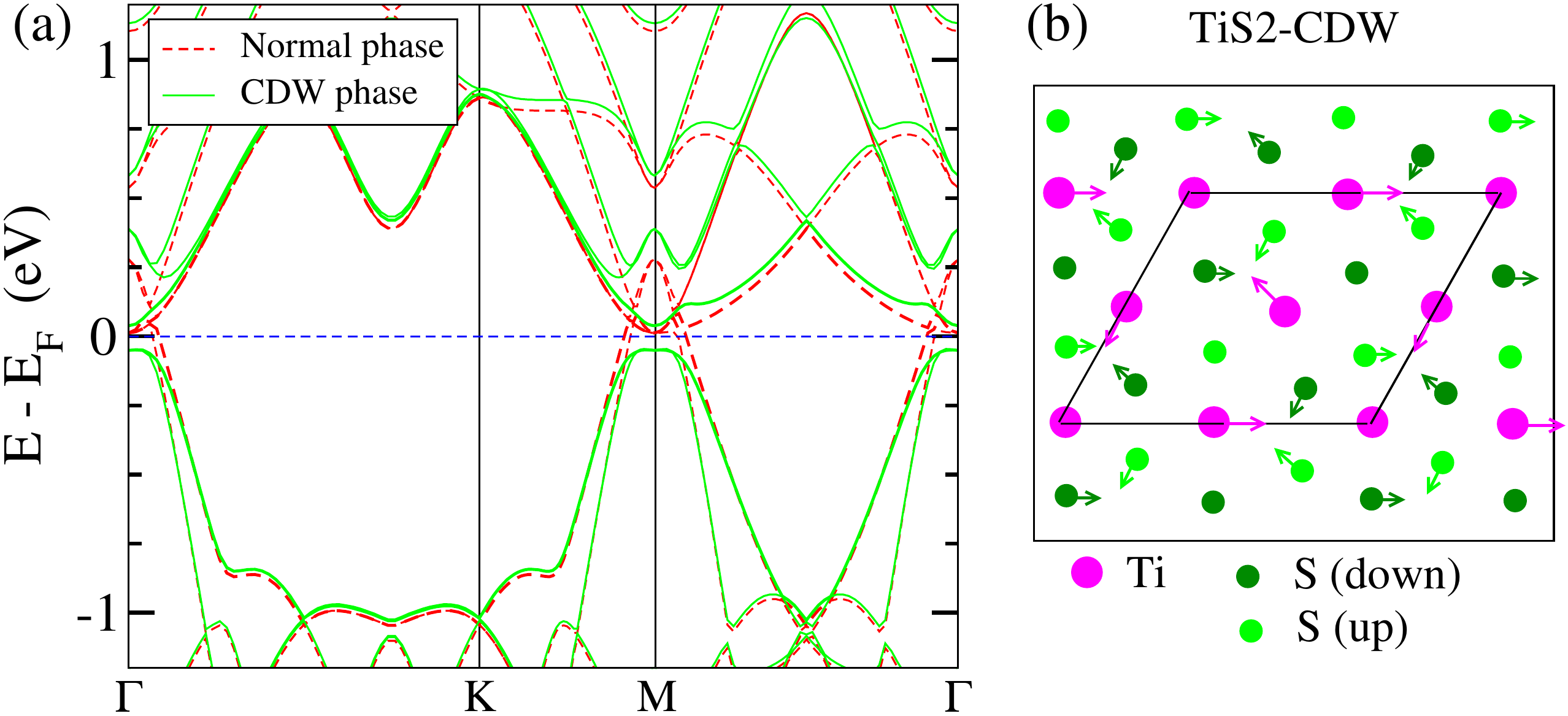}
\caption[]{(Color online) (a) The LDA bandstructure of a TiS$_2$ monolayer 2$\times$2 supercell 
in the normal (red dashed lines) and the commensurate CDW (green lines) phases. (b) The cartoon 
shows the commensurate CDW phase of the TiS$_2$ monolayer (top view). The arrows indicate the 
planar displacement of the various atoms.}
\label{fig:Bands_schematic}
\end{figure}

The phonon dispersion of TiS$_2$ in the monolayer form is shown in Fig.~\ref{fig:PH_pristine}(b). Similarly to 
the bulk phonon dispersion also for the monolayer there is no energy-gap between the optical and the acoustic 
phonon modes. However, we find that when TiS$_2$ is thinned down to a monolayer the Kohn instability at M
becomes a dynamically instability, namely there is now an imaginary phonon frequency at M. This reflects the fact 
that the system gets stabilized by forming a 2$\times$2 superstructure, i.e. a commensurate CDW phase where
the atomic positions are distorted. If one now constructs a 2$\times$2 supercell such an instability migrates at the 
$\Gamma$ point, reflecting the fact that M folds to $\Gamma$ in the supercell. Now atomic relaxation of the supercell
gives us a new distorted structure, which is 1.1~meV/formula-unit lower in energy than the undistorted one. The 
phonon bandstructure for such new distorted phase has now only real frequencies, indicating that no further
symmetry lowering is present. A schematic diagram of the distorted superstructure is shown in 
Fig.~\ref{fig:Bands_schematic}(b). The direction of the distortion of the atomic positions are determined by the 
$q$-vector, at which the unstable phonon mode occurs. In such new CDW phase the displacement of the Ti 
atoms, $\delta\zeta_{Ti}$ is much larger, 0.05 \AA, than that of S~$\leq$~0.005 \AA. 

The bandstructure for the monolayer as calculated for the 2$\times$2 supercell is presented in 
Fig.~\ref{fig:Bands_schematic}(a) for both the normal and the CDW phase. The crucial aspect is that
in the CDW phase the material is able to open a bandgap of 0.09~eV, i.e., as expected, the emergence 
of the CDW phase is associated to a metal to insulator transition. The microscopic origin of the CDW 
phase is mainly associated to the strong electron-phonon coupling in LTMDs, which allows one to modify 
significantly the bandstructure with moderate lattice distortions. This is a feature common to the entire
class of LTMD materials~\cite{JPSJ_1994_63_156, PRL_1975_32_117}. A similar CDW phase transition 
has been observed experimentally in bulk TiSe$_2$, a compound iso-structural to TiS$_2$, and the transition 
temperature was reported to be as large as 200 K~\cite{NL_2012_12_5941}. 

Intriguingly, we find that the formation of the CDW phase is not a unique feature of the TiS$_2$ monolayer. 
In fact, we have repeated the phonon calculation for the TiS$_2$ unit cell and a number of layers ranging 
from one to four and found imaginary phonon frequencies in all cases. In particular we report frequencies of
-92~cm$^{-1}$, -52~cm$^{-1}$, -18~cm$^{-1}$ and -21~cm$^{-1}$, respectively for one, two, three and four layers 
(note that here we report the purely imaginary part of the frequency). Considering that our accuracy over the 
frequency is of the order of 20~cm$^{-1}$, we can conclude that the CDW phase is certainly present for both 
mono- and bi-layers and it vanishes for structures containing between three and four layers. This indicates
that the interlayer interaction plays an important role in determining the condition for the CDW instability.
  
\begin{figure}[ht]
\center
\includegraphics[width=8.0cm,clip=true]{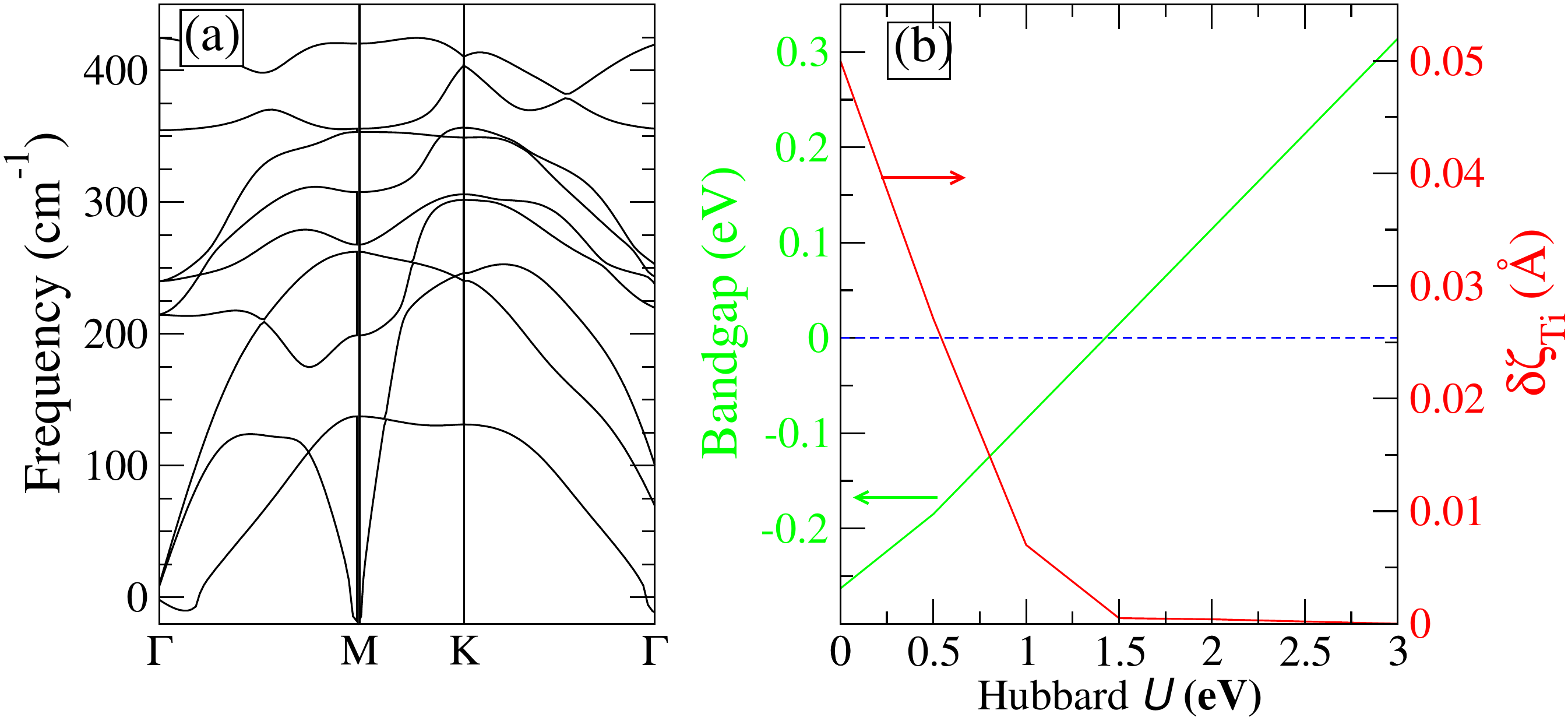}
\caption[]{(Color online) (a) Phonon dispersion of a TiS$_2$ monolayer under a compressive strain of 6.6~\%. 
(b) Variation of the TiS$_2$ monolayer bandgap (green) of the undistorted phase and the Ti-distortion, $\delta \zeta_{Ti}$ (red), 
against the Hubbard $U$ parameter for LDA+$U$ calculations. A negative bandgap indicates a semi-metallic ground
state.}
\label{fig:PH_LDAU}
\end{figure}

Another possibility for suppressing the CDW phase is by applying isotropic pressure, as already suggested for
bulk TiSe$_2$~\cite{PRL_2011_106_196406}. This can be simulated by performing phonon bandstructure
calculations for isotropically compressed unit cells and here we report results for the monolayer case. Importantly 
such compression does not alter significantly the electronic structure, and the material remains metallic at
all the pressures investigated with the only notable effect being an increase of the S-$p$ bandwidth as the pressure 
gets larger. The situation, however, is different for the phonon bandstructure. As a demonstration in 
Fig.~\ref{fig:PH_LDAU}(a) we present data for a compressive strain of 6.6\%. Interestingly, the minimum frequency 
of the soft-phonon mode at the M point has an imaginary value (-8~cm$^{-1}$) significantly smaller than the case
when no pressure is applied (-92~cm$^{-1}$). Consequently, the energy gain due to the CDW formation reduces 
with increasing the compressive strain. Eventually, the unstable mode disappears for a compressive strain of 7.1\%, 
which corresponds to a hydro-static pressure of $\approx$~3.9~GPa. Such disappearance of the CDW can be 
explained by considering how the inter-atomic force constants change under compressive strain. Increasing the 
pressure causes the Ti-S bond length to decrease and enhances the stiffness of the nearest neighbors Ti-S force 
constant. As a consequence the energy of the soft-phonon mode increases and eventually the instability is removed. 
As such the CDW instability is sensitive on the local environment around the transition metal atoms in the 
TiS$_2$ layer.

As mentioned before and demonstrated in Fig.~\ref{fig:Bands_schematic} the formation of the CDW is associated
to a metal to insulator transition. One then expects the CDW phase to emerge only in materials that are either metals 
or small-bandgap semiconductors in their undistorted geometry. The case of TiS$_2$ is somehow unclear since there
exists significant ambiguity on the experimental determination of the bandgap (or its absence). From the theoretical
point of view it is important to mention that the LDA systematically underestimates the bandgap, so that the calculated 
semi-metal ground state might not correspond to the correct one. To complicate the comparison with experiments there
is also the fact that in LTMD monolayers the exciton binding energy can be as large as 1~eV, meaning that the 
quasi-particle and the optical gap may differ significantly from each other~\cite{PRB_86_241201}. As such it is useful 
to revisit our results this time by using an exchange and correlation functional, which allows us to systematically open 
a bandgap in the undistorted phase. Considering the predominant 3$d$ character of the valence band, the 
LDA+$U$ functional~\cite{LDAU} appears to be an ideal choice, since the Coulomb repulsion $U$ parameter, can 
be used a control parameter for the opening of the gap. 

In Fig.\ref{fig:PH_LDAU}(b) we report our results showing the variation of the bandgap of the undistorted phase and 
the Ti displacement, $\delta \zeta_{Ti}$, as a function of $U$. Clearly we observe that the lattice distortion vanishes as 
soon as $U$ is large enough to open a bandgap in the undistorted phase. In other words the metallicity on the undistorted
phase is a necessary condition for the emerge of the CDW ground state. At this point we are not in the position to make
a final judgment on the metallicity of TiS$_2$, however our results show convincingly that the electron phonon coupling
in TiS$_2$ is strong enough to drive a CDW instability as long as the material is metallic. This means that, even if
pristine TiS$_2$ is a small-gap insulator, small doping levels (e.g. by intercalating alkali atoms) will be sufficient to 
trigger the instability. 

%CONCLUSION
%**********************************************************************
%\section{Conclusion}
%**********************************************************************
{\it{Conclusion}}. In summary, we have performed electronic and phononic bandstructure calculations for TiS$_2$
both in the bulk and monolayer form. In both cases the electronic structure derived from DFT-LDA is that of a semimetal
with the band overlap points being different depending on the dimensionality. More interestingly we have found that,
in contrast to other Ti-based chalcogenides, the bulk form of TiS$_2$ is undistorted, while a CDW instability emerges
for nanostructures made of less than four monolayers. Furthermore the CDW instability can be controlled by pressure
and for the monolayer it vanishes at around 4~GPa. Since the emergence of the CDW phase depends of the metallic 
state of the undistorted structure we have discussed the issue of the bandgap. Our results point to a situation where the 
material can distort either in its pristine form or in condition of moderate doping.

{\it{Acknowledgement}}. This work is supported by CRANN. We thank Trinity Centre for High Performance 
Computing (TCHPC) for the computational resources provided. 
%**********************************************************************
%BIBLIOGRAPHY
%**********************************************************************


\begin{thebibliography}{100}
%\bibitem{PRB_1980_21_615}C. H. Chen, W. Fabian, F. C. Brown, K. C. Woo, B. Davies, B. DeLong and A. H. Thompson, Phys. Rev. B {\bf 21}, 615 (1980).
\bibitem{PRB_1980_21_615}C. H. Chen {\it et al.}, Phys. Rev. B {\bf 21}, 615 (1980).

\bibitem{JPChS_1965_26_1445}D. L. Greenaway and R. Nitsche, J. Phys. Chem. Solids {\bf 26}, 1445 (1965).

\bibitem{JPCssp_1984_17_2713}P. C. Klipstein and R. H. Friend, J . Phys. C : Solid State Phys. {\bf 17}, 2713 (1984).

\bibitem{JPC_1977_10_705}R.H. Friend, D. Jerome, W.Y. Liang, C. Mikkelsen and A.D. Yoffe, J. Phys. C {\bf 10}, L705 (1977).

\bibitem{JPC_1983_16_393}J.J. Barry, H.P. Hughes, P.C. Klipstein and R.H. Friend, J. Phys. C {\bf 16}, 393 (1983).

\bibitem{PRL_1972_29_163}A.H. Thompson, K.R. Pisharody and R.F. Koehler, Phys. Rev. Lett. {\bf 29}, 163 (1972).

\bibitem{PRB_1973_7_3859}G. Lucovsky, R.M. White, J.A. Benda and J.F. Revelli, Phys. Rev. B {\bf 7}, 3859 (1973).

\bibitem{PRB_1973_8_3576}D.W. Fischer, Phys. Rev. B {\bf 8}, 3576 (1973).

\bibitem{AdvP_1987_36_1}R.H. Friend and A.D. Yoffe, Adv. Phys. {\bf 36}, 1 (1987).

\bibitem{Sc_1976_192_1126}M.S. Whittingham, Science {\bf 192}, 1126 (1976).

\bibitem{AdvP_1989_38_565}M. Inoue, H.P. Hughes and A.D. Yoffe, Adv. Phys. {\bf 38}, 565 (1989).

%\bibitem{JAP_2011_109_053717}B. Liu, J. Yang, Y. Han, T. Hu, W. Ren, C. Liu, Y. Ma, and C. Gao, J. Appl. Phys. {\bf 109}, 053717 (2011).

\bibitem{JAP_2011_109_053717}B. Liu {\it et al.}, J. Appl. Phys. {\bf 109}, 053717 (2011).

\bibitem{JPCM_2011_23_055401}Y. G. Yu and N. L. Ross, J. Phys.: Condens. Matter {\bf 23}, 055401 (2011).

\bibitem{JPCssp_1985_18_1595}G. . Benesh, A.M. Woolley and C. Umrigar, J . Phys. C: Solid State Phys. {\bf 18}, 1595 (1985).

\bibitem{SSSc_2010_12_1786}F. Yu, J.-X. Sun and Y.-H. Zhou, Solid State Science {\bf 12}, 1786 (2010).

%\bibitem{NJP_2008_10_125027}D. S. Inosov, V B Zabolotnyy , D. V. Evtushinsky, A. A. Kordyuk, B. B\"uchner, R Follath, H Berger and S. V. Borisenko, New J. Phys.~{\bf 10}, 125027 (2008). 

\bibitem{NJP_2008_10_125027}D. S. Inosov {\it et al.},  New J. Phys.~{\bf 10}, 125027 (2008). 

\bibitem{NatNano_2012_7_699}Q. H. Wang, K. Kalantar-Zadeh, A. Kis, J. N. Coleman and M. S. Strano, Nature Nanotech. {\bf 7}, 699 (2012). 

\bibitem{JCScience}V.~Nicolosi, M.~Chhowalla. M.G.~Kanatzidis, M.S.~Strano and J.N.~Coleman, Science 
{\bf 340}, 1420 (2013).

%\bibitem{NL_2010_10_1271}A. Splendiani, L. Sun, Y. Zhang, T. Li, J. Kim, C.-Y. Chim, G. Galli and F. Wang, Nano Lett. {\bf 10}, 1271 (2010). 

\bibitem{NL_2010_10_1271}A. Splendiani {\it et al.}, Nano Lett. {\bf 10}, 1271 (2010). 

\bibitem{NL_2012_12_5941}P. Goli, J. Khan, D. Wickramaratne, R.K. Lake and A. A. Balandin, Nano Lett. {\bf 12}, 5941 (2012). 

\bibitem{PR_136_B864}P.~Hohenberg and W.~Kohn, Phys. Rev. {\bf 136}, B864 (1964).

\bibitem{PR_140_A1133}W.~Kohn and L.J. Sham, Phys. Rev.~{\bf 140}, A1133 (1965).

\bibitem{RMP_73_515}S. Baroni, S. de Gironcoli, A. Dal Corso, and P. Giannozzi, Rev.~Mod.~Phys.~{\bf 73},~515~(2001) .

\bibitem{PRL_1980_45}D.M.~Ceperley and B.J.~Alder, Phys. Rev. Lett.~{\bf 45}, 566 (1980).

\bibitem{QE}P. Giannozzi {\it et al.}, J. Phys.: Condens. Matter {\bf 21}, 395502 (2009).

\bibitem{LDAU}A.I. Liechtenstein, V.I. Anisimov and J. Zaanen, Phys. Rev. B~{\bf 52}, R5467(1995).

\bibitem{PRB_57_5106}D.R. Allan, A.A. Kelsey, S.J. Clark, R. J. Angel and G.J. Ackland, Phys. Rev. B {\bf 57}, 5106 (1998)

\bibitem{AdvP_1969_18_193}J.A. Wilsona and A.D. Yoffe, Adv. Phys. {\bf 18}, 193 (1969).

%\bibitem{ACIE_2011_50_11093}Z. Zeng, Z. Yin, X. Huang, H. Li, Q. He, G. Lu, F. Boey and H. Zhang, Angew. Chem. Int. Ed.~{\bf 50}, 11093 (2011).

\bibitem{ACIE_2011_50_11093}Z. Zeng {\it et al.}, Angew. Chem. Int. Ed.~{\bf 50}, 11093 (2011).

\bibitem{JPCC_116_8983}C. Ataca, H. \c{S}ahin and S. Ciraci, J. Phys. Chem. C {\bf 116}, 8983 (2012).

\bibitem{PRB_1992_45_14347}S.J. Sandoval, X. K. Chen and J.C. Irwin, Phys. Rev. B {\bf 45}, 14347 (1992).

\bibitem{PRB_1986_33_4317}M. Scharli and F. L\'evy, Phys. Rev. B {\bf 33}, 4317 (1986).

\bibitem{JPSJ_1994_63_156}Y. Nishio, M. Shirai, Naoshi, and K. Motizuki, J. Phys. Soc. Jap. {\bf 63}, 156 (1994).

\bibitem{PRL_1975_32_117}J. A. Wilson, F. J. DiSalvo and S. Mahajan, Phys. Rev. Lett. {\bf 32}, 117 (1975).

\bibitem{PRL_2011_106_196406}M. Calandra and F. Mauri, Phys. Rev. Lett. {\bf 106}, 196406 (2011).

\bibitem{PRB_86_241201}H.-P. Komsa and A. V. Krasheninnikov, Phys. Rev. B~{\bf 86}, 241201(R) (2012).
\end{thebibliography}
\end{document}